\documentclass[aps,prf,onecolumn,floatfix,superscriptaddress]{revtex4}
\usepackage{graphicx}
\usepackage{epstopdf}
\usepackage{float}
\usepackage{amsmath}
\usepackage{relsize}
\usepackage{color}
\usepackage{hyperref}
\usepackage{breakurl}

\begin{document}

\title{Turbulent thermal superstructures in Rayleigh-B\'enard convection}
\author{Richard J.A.M. Stevens}
\affiliation{Physics of Fluids Group, Max Planck Center Twente for Complex Fluid Dynamics, J. M. Burgers Center for Fluid Dynamics, and MESA+ Research Institute, University of Twente, P.O. Box 217, 7500 AE Enschede, The Netherlands}
\author{Alexander Blass}
\affiliation{Physics of Fluids Group, Max Planck Center Twente for Complex Fluid Dynamics, J. M. Burgers Center for Fluid Dynamics, and MESA+ Research Institute, University of Twente, P.O. Box 217, 7500 AE Enschede, The Netherlands}
\author{Xiaojue Zhu}
\affiliation{Physics of Fluids Group, Max Planck Center Twente for Complex Fluid Dynamics, J. M. Burgers Center for Fluid Dynamics, and MESA+ Research Institute, University of Twente, P.O. Box 217, 7500 AE Enschede, The Netherlands}
\author{Roberto Verzicco}
\affiliation{Physics of Fluids Group, Max Planck Center Twente for Complex Fluid Dynamics, J. M. Burgers Center for Fluid Dynamics, and MESA+ Research Institute, University of Twente, P.O. Box 217, 7500 AE Enschede, The Netherlands}
\affiliation{Dipartimento di Ingegneria Industriale, University of Rome ``Tor Vergata''. Via del Politecnico 1, Roma 00133, Italy}
\author{Detlef Lohse}
\affiliation{Physics of Fluids Group, Max Planck Center Twente for Complex Fluid Dynamics, J. M. Burgers Center for Fluid Dynamics, and MESA+ Research Institute, University of Twente, P.O. Box 217, 7500 AE Enschede, The Netherlands}
\affiliation{Max Planck Institute for Dynamics and Self-Organization, Am Fassberg 17, 37077 G\"ottingen, Germany}
\date{\today}

\begin{abstract}
We report the observation of superstructures, i.e.\ very large-scale and long living coherent structures in highly turbulent Rayleigh-B\'enard convection up to Rayleigh $Ra=10^9$. We perform direct numerical simulations in horizontally periodic domains with aspect ratios up to $\Gamma=128$. In the considered $Ra$ number regime the thermal superstructures have a horizontal extend of six to seven times the height of the domain and their size is independent of $Ra$. Many laboratory experiments and numerical simulations have focused on small aspect ratio cells in order to achieve the highest possible $Ra$. However, here we show that for very high $Ra$ integral quantities such as the Nusselt number and volume averaged Reynolds number only converge to the large aspect ratio limit around $\Gamma \approx 4$, while horizontally averaged statistics such as standard deviation and kurtosis converge around $\Gamma \approx 8$, and the integral scale converges around $\Gamma \approx 32$, and the peak position of the temperature variance and turbulent kinetic energy spectra only around $\Gamma \approx 64$.
\end{abstract}
\maketitle

\begin{figure*}
\includegraphics[width=0.99\columnwidth]{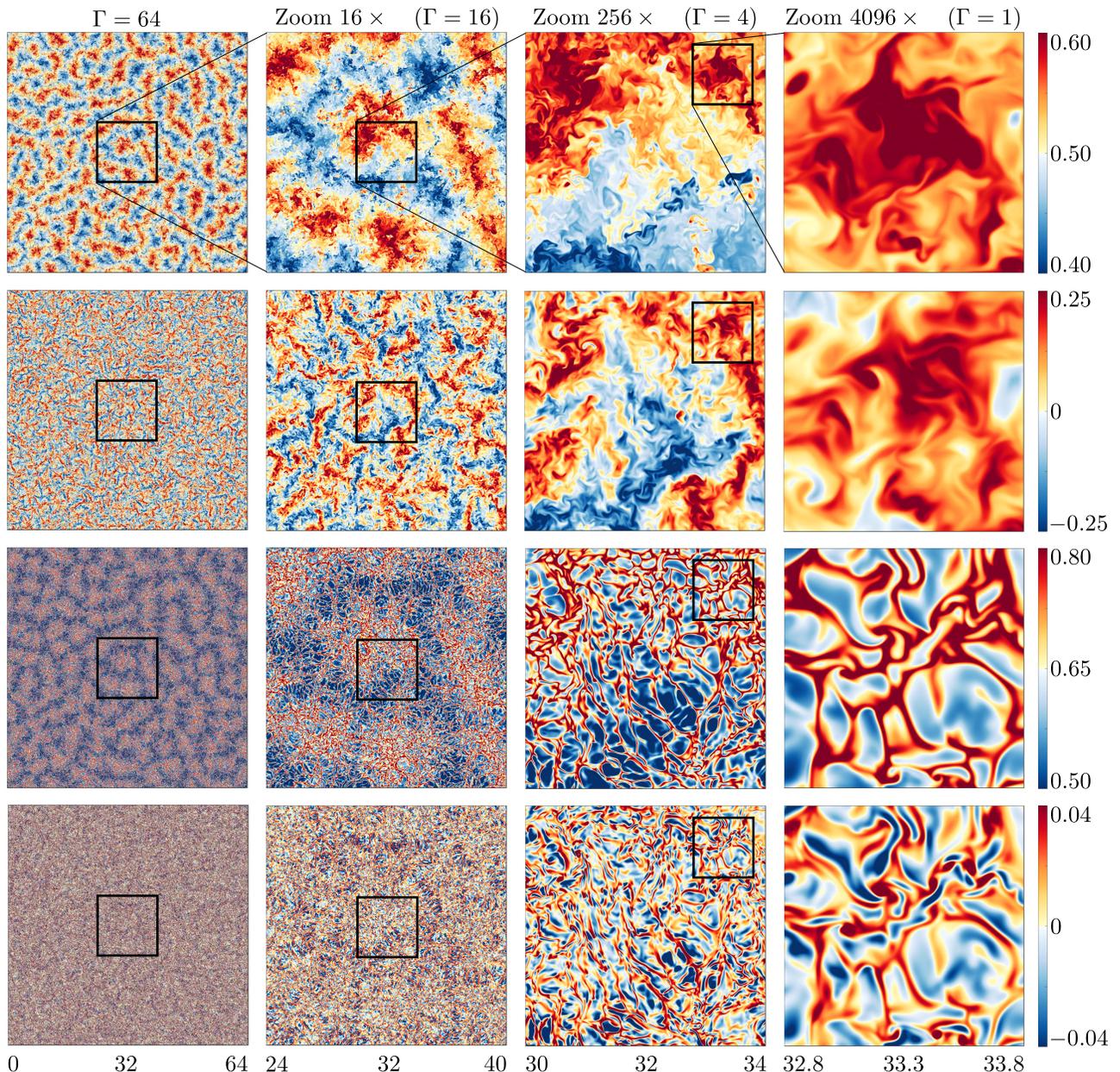}
\caption{Snapshots at different magnifications for the simulation at $ Ra=10^8 $ and $\Gamma=64$. The first and second row show snapshots of the temperature and vertical velocity at mid-height and the third and fourth row show the corresponding snapshots at BL height. The columns from left to right show successive zooms of the area indicated in the black box. }
\label{fig1} 
\end{figure*}

Turbulence is characterized by chaotic, vigorous fluctuations. Therefore it is surprising to observe very large-scale coherent structures in turbulent flows such as channel \cite{kim87,lee15}, pipe \cite{eck07}, or turbulent boundary layer flows \cite{mar10,smi11,jim12}.\ To observe these superstructures [figure \ref{fig1}], very large experimental or numerical domains are necessary. So far, superstructures have been observed in pressure and shear driven turbulent flows. However, up to now they have not been reported in highly turbulent thermally driven turbulence, where a preferred flow direction is absent, reflected in the community's focus on experiments and simulations in small aspect ratio cells. Here we study thermal superstructures, defined as the largest horizontal flow scales that develop, such that their flow characteristics, size, and shape are independent of the system geometry, in highly turbulent thermal convection. So, even though the large-scale circulation (LSC) observed at very high $Ra$ in $\Gamma\sim0.5-1.0$ cells is a fascinating feature of flow organization \cite{he12a,ahl12b,he15,ahl09,chi12}, such a LSC in a confined cell is not a thermal superstructure according to our definition since the geometrical and dynamical features depend on the system geometry.

The most popular model of thermal convection is RB flow \cite{kad01,ahl09,loh10,chi12,xia13}, where the dimensionless control parameters are the Rayleigh ($Ra$) and Prandtl ($Pr$) numbers, parameterizing the dimensionless temperature difference and the fluid properties. Major advances have been achieved in the last few decades in theoretically understanding the global transfer properties of the flow. Namely, the unifying theory of Refs.\ \cite{gro00,gro01,ste13} describes the Nusselt $Nu$ (dimensionless heat transport) and Reynolds $Re$ number (dimensionless flow strength) dependence on $Ra$ and $Pr$ well. In addition, experiments and simulations agree excellently up to $Ra\sim10^{11}$ due to major developments in experimental and numerical techniques \cite{ahl09,loh10,chi12,xia13}. 

However, the effect of the third dimensionless quantity, the aspect ratio $\Gamma=W/L$, where $W$ is the cell's width and $L$ its height, is much less understood. According to the classical view, strong turbulence fluctuations at high $Ra$ should ensure that the effect of the geometry is minimal as the entire phase space is explored statistically by the flow \cite{kol41a,kol41c}. This view would justify the use of small aspect ratio domains, which massively reduces the experimental or numerical cost to reach the high $Ra$ number regime relevant for industrial applications and astrophysical and geophysical phenomena, while maintaining the essential physics. Therefore, in a quest to study RB convection at ever increasing $Ra$, most experiments and simulations have focused on relatively small aspect ratio and have been performed for $\Gamma \lesssim2$, while very high $Ra$ number cases are even limited to $\Gamma \sim 0.2-0.5$. This approach allowed the discovery of the ultimate regime of thermal convection \cite{he12}, already predicted by Kraichnan in $1962$ \cite{kra62} and later by Grossmann and Lohse \cite{gro11}. 

However, while heat transfer in industrial applications takes place in confined systems, the aspect ratio in many natural instances of convection is extremely large \cite{ahl09,loh10,chi12,xia13}. Very large flow patterns have, for example, been observed in moist convection simulations \cite{sch10d,wei10b,pau11}, high $Ra$ number non-penetrative convection \cite{adr86}, and in plane Couette \cite{bar05,bar07} at low $Re$. For RB convection just above the onset of convection, experiments \cite{fit76,sun05e,pui07c,xia08,zho12,hog13,pui13} and simulations in large periodic \cite{har03,par04,har05,har08} and in very large aspect cylindrical \cite{shi05,shi06,shi07,bai10,sch14,emr15,sak16} domains have revealed beautiful flow patterns \cite{bod00,mor93,ass93}. Correlations between single point measurements \cite{pui07c,pui13} and PIV measurements at high $Ra$ \cite{xia08} have shown a transition between a single and multi-roll structure when $\Gamma$ exceeds roughly $4$, while simulations at $\Gamma=6.3$ and $Ra=9.6\times10^7$ \cite{sak16} show that large regions of warm rising and cold sinking fluid are still present. Previously, Hartlep {\it et al.} \cite{har03,har05} showed with simulations from the onset up to $Ra=10^7$ and for aspect ratios up to $20$ that the size of the largest flow structures increases with increasing $Ra$. These simulation results agree well with the classical experiments performed by Fitzjarrald \cite{fit76} in rectangular containers. In addition, Hartlep {\it et al.} \cite{har03,har05} showed that at $Ra=10^5$ and $10^6$ the size of thermal superstructures peaks at {\it intermediate} $Pr$. Also Parodi {\it et al.} \cite{par04} showed the emergence of large scale flow patterns up to $Ra=10^7$ and $\Gamma=2\pi$. Later, von Hardenberg {\it et al.} \cite{har08} showed in simulations with aspect ratios up to $12\pi$ that, after an initial growth period, the size of thermal superstructures becomes constant as function of time.

However, there is no clear insight into the development of thermal superstructures at higher $Ra$. In this regime the behavior could be quite different as only for these high $Ra$ the flow becomes so turbulent that the coherence length becomes considerably smaller than $0.1L$ \cite{sug09}. In this previously ``unexplored" highly turbulent regime classical theories would predict that turbulent superstructures disappear. Here we will show (i) that thermal superstructures survive at high $Ra$, (ii) that the thermal superstructures have pronouncedly different flow characteristics than the LSC in smaller domains, and (iii) that the domain size to obtain convergence to the large aspect ratio limit depends on the quantity of interest. 

We performed direct numerical simulations (DNS) of periodic RB convection in very large computational domains and at high $Ra$ using \textsc{AFiD}. \textsc{AFiD} uses a second order, energy conserving, finite difference method. Here, we use no-slip conditions, constant temperature boundary conditions at the bottom and top plates, and periodic boundary conditions in the horizontal directions. Details can be found in Refs.\ \citep{ver96,poe15c,zhu18b} and at www.AFiD.eu. The control parameters are $Ra=\alpha g \Delta L^3/(\nu \kappa)$ and $Pr=\nu/\kappa$, where $\alpha$ is the thermal expansion coefficient, $g$ the gravitational acceleration, $\Delta$ the temperature difference between the top and bottom plates, $L$ the height of the fluid domain, $\nu$ the kinematic viscosity, and $\kappa$ the thermal diffusivity of the fluid. We performed $33$ simulations at $Ra=[2\times10^7,10^8, 10^9]$ in the aspect ratio range $\Gamma=[1-128]$ and $Pr=1$. We took great care to perform all simulations consistently and followed the resolution criteria set in \cite{ste10,shi10}. The simulation at $Ra=10^{9}$ for $\Gamma=32$ is performed on a $12288\times12288\times384$ grid. The statistical convergence of integral flow quantities such as $Nu$ and $Re$ is within a fraction of a percent. The convergence of higher order statistics is, unavoidably, less due to the slow dynamics of the thermal superstructures, whose existence will be revealed. 

We first look at a visualization of the flow at $Ra=10^8 $ in a $ \Gamma=64 $ cell in figure \ref{fig1}. The first row displays the temperature field at mid-height. The different subfigures in this row present the flow structures more clearly by successive zoom-ins. One can easily discern the significance of a sufficiently large aspect ratio. The second row, which shows the mid-height vertical velocity field, displays a remarkable disparity with the temperature field. We find that in large aspect ratio cells the correlation coefficient between temperature and vertical velocity is only about $0.5$ at mid-height, while this correlation is about $0.7$ at BL height. The third and fourth row show the temperature and vertical velocity at BL height.\ It is impressive to see that the large-scale thermal structures at mid-height leave a visible imprint in the BL, i.e.\ the warm (red) areas at mid-height (top row) correspond to warm areas in the BL (third row). This imprint is quantified by the correlation of the temperature field at mid-height and BL height, which is about $0.3$, i.e.\ small but statistically relevant.

\begin{figure*}[t]
\includegraphics[width=\columnwidth]{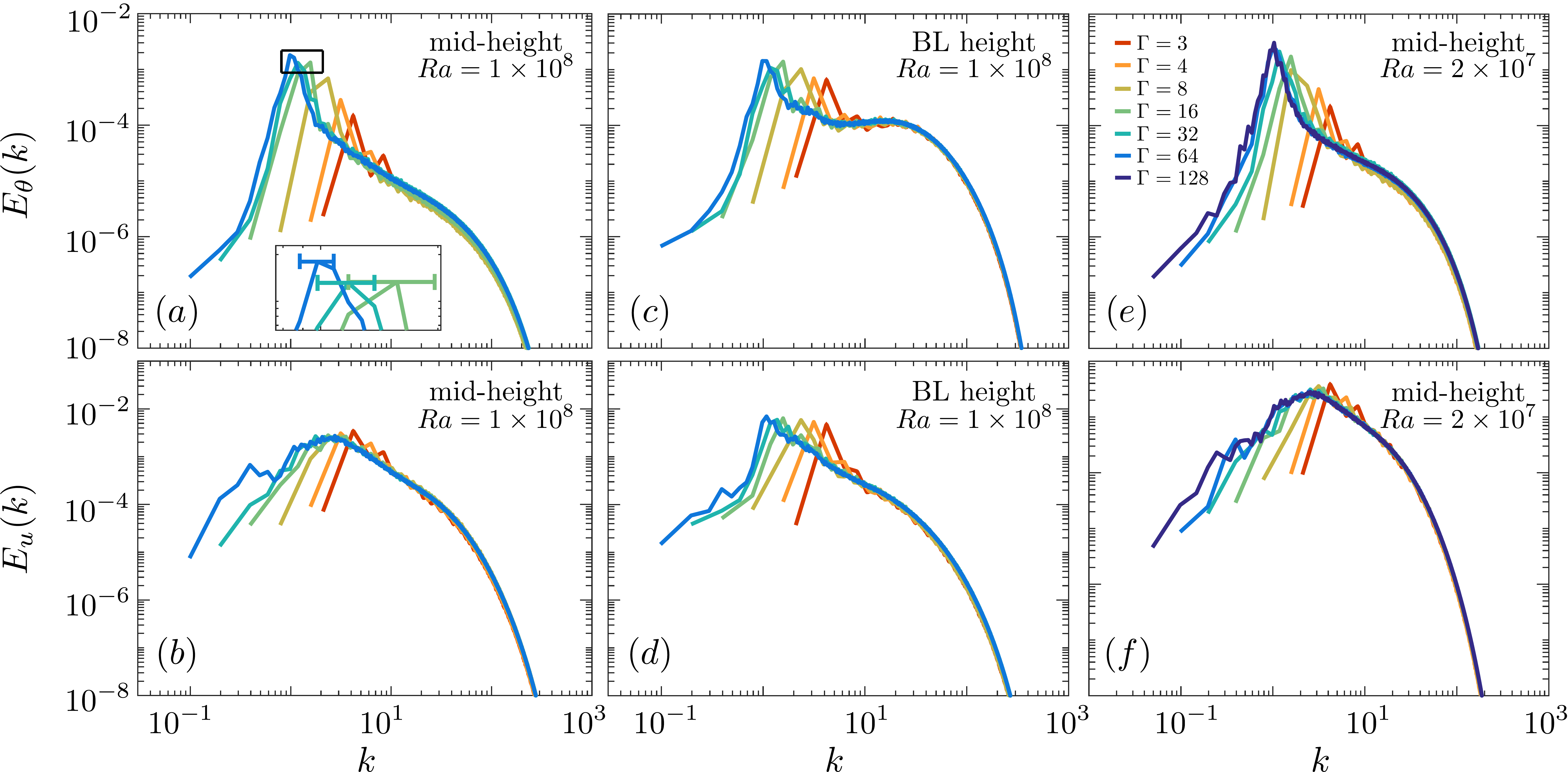}
\caption{The temperature variance $E_\theta(k)$ and TKE $E_u(k)$ spectra at (a,b) mid-height and (c,d) BL height at $ Ra=10^8 $ and (e,f) at mid-height at $ Ra=2\times 10^7$ for different $\Gamma$. Here $ k $ is the circular wave number $ k=(k_x^2+k_y^2)^{1/2} $. The zoom in panel a shows the peaks of $ \Gamma=16,32,64 $ with error bars displaying the distance to the next captured wavenumber.}
\label{fig4} 
\end{figure*}

To determine the horizontal extent of the thermal superstructures, we calculated the turbulent kinetic energy (TKE) $E_u(k)$ and the thermal variance $E_\theta(k)$ spectra at BL height and mid-height.\ The spectra represent the time average obtained in the statistical stationary state. Figure \ref{fig4} shows that the wavenumber of maximal energy, respectively, thermal variance decreases with increasing aspect ratio until it slowly saturates, but for $Ra=10^8$ we cannot conclude whether the peak position of the spectra is fully converged. Figure \ref{fig4}e,f shows that the results for $Ra=2\times10^7$ do reveal a clear convergence of the peak location of the spectra around $\Gamma\approx64$. The slow convergence of the peak location of the spectra shows that extremely large domains are necessary to accurately capture thermal superstructures, i.e.\ the domain size must be much larger than the average size of the superstructures. The temperature variance spectra at mid-height and BL height show that the spectrum peak is located around $k\approx1$, which corresponds to a structure size of about $6-7$ times the domain height. This size is similar as obtained in the classical works  \cite{har03,par04,har05,har08} for $Ra$ up to $10^7$. Also the peak of the TKE spectrum at BL height is located around $k\approx1$. This reflects the large-scale pattern visible in the horizontal velocity components as the spectrum of the vertical velocity component at BL height peaks around $k\approx30$ ($\approx0.21L$), which corresponds to the plume size at BL height shown in figure \ref{fig1}, see also Ref.\ \cite{par04}. For $Ra=10^8$ and $\Gamma=32$ we verified that the shown spectra are converged up to $k\approx700$ by performing separate simulations on a $6144\times 6144\times 192$ and a $8192\times 8192\times 256$ grid. For the TKE spectrum at mid-height the main peak is located close to $ k=2 $, which indicates that the velocity structures at mid-height are smaller than the temperature structures. This further emphasizes that the correlation between the vertical velocity and temperature at mid-height is less than naively expected.

As the location of the peak of the spectrum is difficult to converge we look at the so-called integral length scale \cite{par04}. Here we calculate the integral length scale based on the temperature variance $ \Lambda_\theta= 2\pi \int [E_\theta(k)/k]dk / \int E_\theta(k) dk $ and TKE $ \Lambda_u= 2\pi \int [E_u(k)/k]dk / \int E_u(k) dk $ spectra. We emphasize that the integral length scales do not correspond to the spectral peaks discussed in the previous section. Figure \ref{fig2}a reveals that $\Lambda_\theta$ and $\Lambda_u$ converge to a large aspect ratio limit around $\Gamma \approx 32$. For $\Gamma \lesssim 8$, when there is only one  convection roll in the domain, $\Lambda_\theta$ and $\Lambda_u$ increase roughly linearly with the domain size. For $\Gamma=16$ and low $Ra$ $\Lambda_\theta$ is similar to the value found for $\Gamma=8$, while for higher $Ra$ it is close to the large aspect ratio limit. We speculate that this phenomenon could be due to the existence of multiple turbulent states at $\Gamma=16$, similarly to what is observed in, for example, Taylor-Couette \cite{hui14} and 2D RB flow \cite{poe11,poe12}.

\begin{figure*}[t]
\includegraphics[width=\columnwidth]{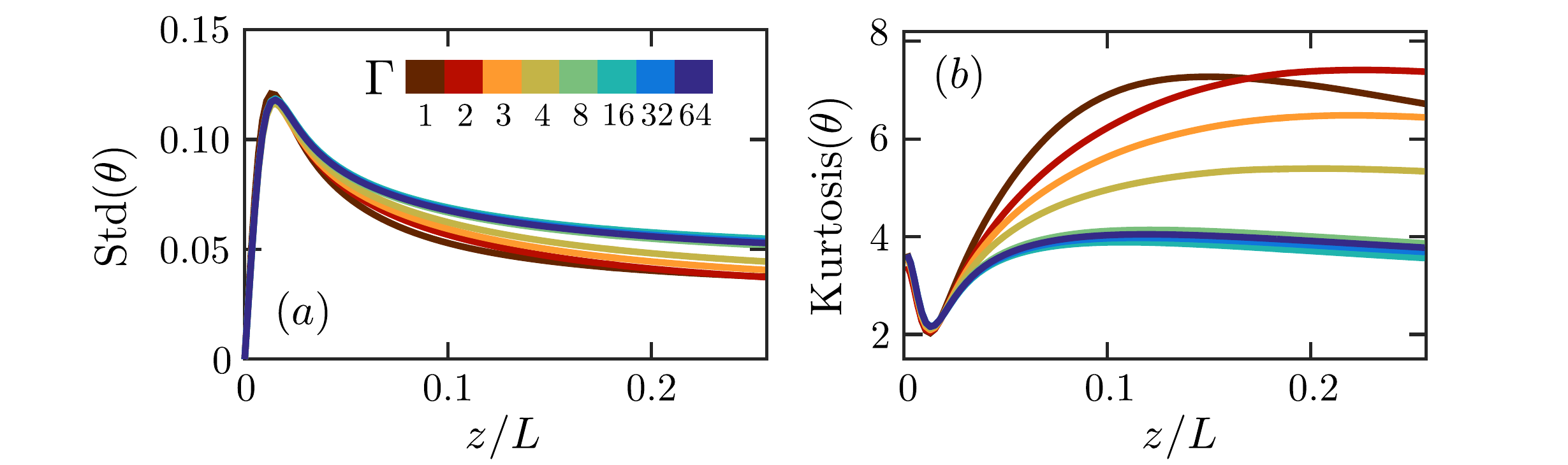}
\caption{Height profile of (a) the standard deviation and (b) the kurtosis at $Ra=10^8$.}
\label{fig5}
\end{figure*}

To investigate how the flow structures influence horizontally averaged higher order statistics, we show the standard deviation and kurtosis of the temperature as function of height in figure \ref{fig5}. The curves show a clear separation between the flow characteristics in small and in large aspect ratio cells. In contrast to the spectra and integral length scales we find that horizontally averaged higher order statistics already converge to the large aspect ratio limit around $\Gamma\approx8$, while figure \ref{fig2} reveals that integral quantities such as $Nu$ and $Re$ are already converged around $ \Gamma \approx 4 $. Figure \ref{fig2}b shows that $Nu$ as function of $\Gamma$ reaches a maximum around $ \Gamma \approx 0.75 $ for all $Ra$ considered here, while it decreases sharply for $\Gamma\lesssim0.5$. The figure also reveals that in smaller domains the horizontal motion is suppressed, while the vertical motion and heat transfer in the system are much less sensitive to the aspect ratio. Consequently, the vertical velocity is much stronger than the horizontal velocity in small domains, while the horizontal and vertical velocity components are nearly equal in large aspect ratio domains, see figure \ref{fig2}e. Thus in large aspect ratio cells the horizontal mixing in the interior domain is stronger than in smaller aspect ratio cells, which results in the lower correlation of the temperature and vertical velocity observed in large aspect ratio cells. We find that the correlation between temperature and vertical velocity converges to the large aspect ratio limit around $\Gamma=8$.

In summary, we highlighted the existence of thermal superstructures in highly turbulent RB flow. The observed structure sizes are significantly bigger than the structures found near the onset of convection \cite{bod00} or the structures found in 2D RB \cite{poe11,poe12}, but similar in size as obtained in  the classical works \cite{har03,par04,har05,har08}, which studied thermal superstructures up to $Ra=10^7$ in simulations with aspect ratio up to about $40$. Surprisingly, while classical theory would predict that these flow structures should be washed out at high $Ra$ when the flow becomes  turbulent, we do not find any sign that the superstructures get weaker when $Ra$ is increased. Our simulations show for the first time that the peak location of the temperature variance and TKE spectra converge around $\Gamma\approx64$, which shows that the characteristics of the thermal superstructures become truly independent of the domain size. Here we also show that the horizontal velocity increases rapidly when the domain size is enlarged until it converges to its large aspect ratio limit around $\Gamma \approx 4$. This leads to more vigorous mixing in large domains, which is reflected in the lower correlation between temperature and vertical velocity in large domains when compared to small domains. While the vertical velocity and heat transfer are much less sensitive to the domain size, we find that the large-scale motions have a visible effect on the heat transfer.

\begin{figure*}[t]
\includegraphics[width=\columnwidth]{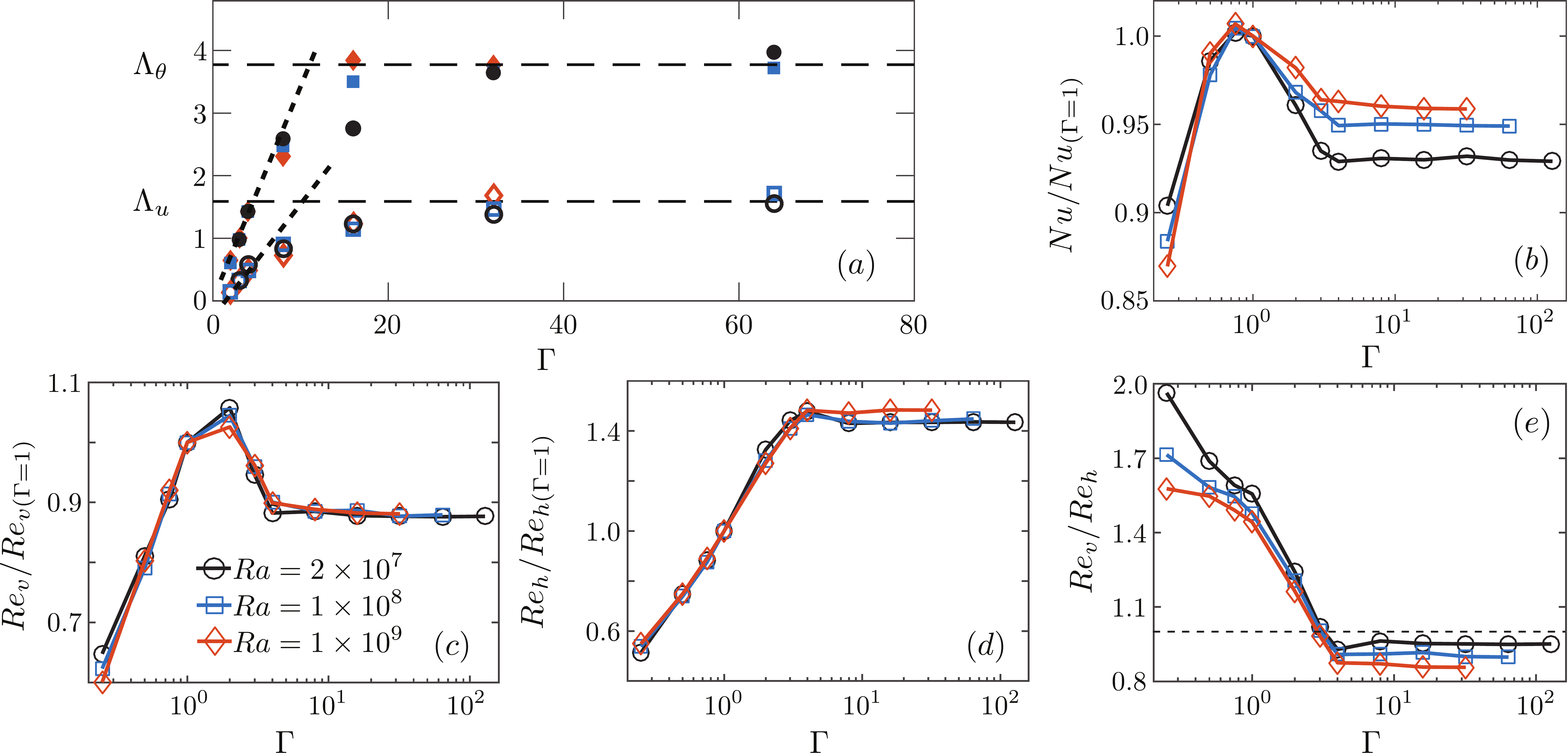}
\caption{(a) Integral scale at mid-height based on the temperature variance $ \Lambda_\theta $  and the TKE spectra $ \Lambda_u $, (b) Nusselt number, (c) vertical Reynolds number $Re_v$, and (d) horizontal Reynolds number $Re_h$ as function of $\Gamma$ normalized with the value at $\Gamma=1$. (e) $Re_v/Re_h$ as function of $\Gamma$.}
\label{fig2} 
\end{figure*}

Thermal superstructures have a profound influence on flow statistics. Interestingly, we find that integral quantities such as $Nu$ and $Re$ reach the large aspect ratio limit already around $\Gamma \approx 4$,\ while this limit is only reached around $\Gamma \approx 8$ for horizontally averaged higher order statistics, around $\Gamma \approx 32$ for the integral length scales, and around $\Gamma\approx64$ for the peak location of the temperature variance and TKE spectra. Thus the minimal domain size required to reach the large aspect ratio limit result depends on the quantity of interest.\ The observation that simple statistics are accurately captured in a smaller domain than necessary to converge spectra is similar to the situation in channel \cite{kim87,lee15}, pipe \cite{eck07} and turbulent boundary layer flow \cite{mar10,smi11,jim12}.\ However, we note that the thermal superstructures are very different than large-scale structures discovered in pipe, channel, and boundary layer flow. First of all, the absence of a mean flow direction means that the thermal superstructures have a similar extend in all horizontal directions, whereas in channel, pipe and boundary layer flows the large-scale flow features are very long and elongated \cite{mar10,smi11,jim12}. In addition, the thermal superstructures have a size that is independent of the distance to the wall, while superstructures appear to be limited to the logarithmic region in turbulent boundary layer flow \cite{mon09b} or the outer layer for pipe and channel flow \cite{bai10c}. 

Further research is required to investigate how the $Pr$ number influences the formation of thermal superstructures at high $Ra$. In addition, the observations that the kurtosis of the temperature distribution convergence to the Gaussian value, and the weak correlation between the temperature and vertical velocity in the bulk are very intriguing phenomena and need further investigation in order to determine how these observations are related to the coherency of the large-scale flow patterns.

\begin{acknowledgments}
This work is supported by the Dutch Foundation for Fundamental Research on Matter (FOM), and by Netherlands Center for Multiscale Catalytic Energy Conversion (MCEC), both funded by the Netherlands Organization for Scientific Research (NWO). We thank the German Science Foundation (DFG) for support via program SSP 1881. The authors gratefully acknowledge the Gauss Centre for Supercomputing (GCS, {\color{blue}\url{www.gauss-centre.eu}}) for providing computing time for a GCS Large-Scale Project on the GCS share of the supercomputer SuperMUC at Leibniz Supercomputing Centre (LRZ, {\color{blue}\url{www.lrz.de}}) under grant 10628/11695. GCS is the alliance of the three national supercomputing centres HLRS (Universit\"at Stuttgart), JSC (Forschungszentrum J\"ulich), and LRZ (Bayerische Akademie der Wissenschaften), funded by the German Federal Ministry of Education and Research (BMBF) and the German State Ministries for Research of Baden-W\"urttemberg (MWK), Bayern (StMWFK) and Nordrhein-Westfalen (MIWF). Part of the work was carried out on the national e-infrastructure of SURFsara, a subsidiary of SURF cooperation, the collaborative ICT organization for Dutch education and research.
\end{acknowledgments}

\end{document}